\relax
\documentclass[12pt]{article}
\usepackage{latexsym}
\usepackage[]{}

\begin{document}

\title{\large\bf Chaotic Solutions Of The Nonlinear Schr\"odinger 
Equation In Classical And Quantum Systems}

\author{F V Kusmartsev $^{1,2}$, K E K\"urten $^3$ and H S Dhillon$^1$}

\date{}

\maketitle

{\bf  $^1$ Department of Physics, Loughborough University, LE11 3TU, UK}

{\bf $^2$ Landau Institute, Moscow, Russia }

{\bf $^3$ 
 Institut f\"ur Experimentalphysik, Universit\"at Wien, Austria}

\vspace{5mm}
\begin{abstract}

We discuss stationary solutions of the nonlinear Schr\"odinger
equation (NSE) applicable to several quantum spin, electron and
classical lattice systems.  
We show that there may arise chaotic spatial
structures in the form of incommensurate or irregular quantum states
and trajectories in space.

As a first (typical) example we consider a single electron which is 
strongly coupled with phonons
on a 1D chain of atoms. 
In the adiabatic approximation 
the system is conventionally 
described  by a discrete 
set of NSEs.
Another apt example is that of superconducting states
in layered superconductors described by the same
NSE.  Amongst many other applications the typical example for a classical
lattice is a system of coupled nonlinear oscillators.

We reformulate this discrete NSE to the form of a 2D mapping.
By this we may investigate a quantum problem 
by methods conventionally applied to classical 
chaotic dynamics. 
We find three types of solutions: periodic, quasiperiodic
and chaotic.  We then develop 
a procedure which allows us to obtain numerical solutions
of the NSE directly.  This procedure may be used to any arbitrary 
accuracy and so these solutions are exact to the degree of precision 
specified.  Both methods give a consistent result.  When applied to our 
typical example we find that the wave function
of an electron on a deformable lattice (and other quantum or classical
discrete systems) may exhibit incommensurate
and irregular structures.

\end{abstract}


\section{Introduction}  
  
Chaos is an important branch of nonlinear dynamics because 
chaotic behaviour seems to be universal \cite{Hilborn}.  It is 
present in mechanical oscillators, electrical circuits, lasers, 
nonlinear optical systems, chemical reactions, nerve cells, heated 
fluids and weather systems.  Even more importantly, this chaotic 
behaviour shows qualitative and quantitative universal features 
which are independent of the details of the particular system.

In Quantum Mechanics instead of initial conditions we have 
boundary conditions.  It is commonly believed that in quantum systems 
chaotic structures
cannot arise.  However, the question does arise if it would be possible  
to have in quantum systems a situation analogous to
classical chaotic structures?  In 
other words, if it would be possible for two slightly different 
boundary conditions or some physical parameters 
(for example, coupling constant) to correspond to two qualitatively 
different wave 
functions?  Such dependence on physical conditions in quantum
systems may be analogous to classical chaotic dynamics.

Classical chaos is more often seen in discrete systems, 
which can be described, for example, by a discrete map.  
This was first done by the seminal work 
of Feigenbaum with the aid of iterative maps (\cite{Hilborn} and 
references therein) where he showed that for a large class of 
iterative maps, exhibiting 
infinite period-doubling bifurcations, the transition to aperiodic 
orbits can be described by universal constants, governing the 
so-called Feigenbaum route to chaos.  
Classical chaos corresponds to a disappearance of periodic 
trajectories.
As classical chaotic motion is more obvious in time discrete 
systems, it is therefore natural to study the quantum analogy of this 
phenomenon by considering the quantum effects in solids which are 
naturally discrete in space due to their atomic structure.   Consequently, 
we study
the  atomic lattice taking into account a nonlinearity which arises 
due to  electron-phonon interactions.  This gives rise to the creation 
of some self-trapped states.  
Some other systems of nonlinear lattices arising in film deposition may 
also be described as coupled nonlinear oscillators.

\section{The Hamiltonian And The NSE}

The classical and quantum systems mentioned above may be generally
described with the use of the Hamiltonian

\begin{eqnarray}
  H  &=& \sum_{i} | \psi_{i} - \psi_{i+1} |^{2} - \sum_{i} {c \over 2}| 
  \psi_{i} |^{4} - E \sum_{i}е \mid \psi_{i} \mid^{2}е  \label{adpot1}
\end{eqnarray}

\noindent where $\psi_i$ is the wave function of the 
the self-trapped particle on the $i^{th}$ site, $c$ is 
some parameter and $E$ is the energy eigenvalue.  For coupled
(nonlinear) oscillators the function $\psi$ describes a lattice distortion.
The wave function, of course,  must satisfy  the conventional 
normalization condition

\begin{equation}
  \sum_i \mid \psi_{i} \mid^{2} = 1 \label{nc1}
\end{equation}

\noindent which is effectively described when the eigenvalue $E$ is used as a 
Lagrange multiplier. The parameter $c$ may have both positive
(for  self-trapped quantum states) and negative (for nonlinear coupled 
oscillators) values depending on the system being studied.

The equations which describe the stationary critical points of $H$ are 
given by $\hat \nabla H(\psi) \equiv 0$, where $\hat \nabla$ is the 
{\it differential operator} ie ${\partial H \over \partial \psi_{i}е}$.  This gives 
the following discretised nonlinear Schr\"odinger equation (NSE) 
 
\begin{equation}  
   -\psi_{i-1}  + 2 \psi_i - \psi_{i+1} - c|\psi_i|^2 \psi_i  =  
    E\psi_{i}  \label{dns1}
\end{equation}

\noindent For simplicity, only real solutions are discussed and so we make the 
substitution $|\psi_{i}|^2 = \psi_{i}^2$ in the NSE.  If we consider the 
NSE describing nonlinear harmonic oscillators we have 

\begin{equation}  
   -x_{i-1}  + 2 x_{i}е - x_{i+1} + c x_{i}^{3}ее  =  e x_{i}  \label{dns2}
\end{equation}

\noindent which differs from (\ref{dns1}) only by the sign of $c$.   
With the aid of the transformation

\begin{eqnarray}
  x_{n}е &=& (-1)^{n}е \psi_{n}е \\
  e &=& 4 - E
\end{eqnarray}

\noindent we again get 

\begin{equation}  
   -\psi_{i-1}  + 2 \psi_{i} - \psi_{i+1} - c\psi_{i}е^{3}е  =  
    E \psi_{i}  \label{dns4}
\end{equation}

Thus, we find that all solutions and the classification of such solutions 
for the quantum problem associated with the NSE with positive coupling 
constant $(c > 0)$  are also the solutions (and corresponding 
classification) of the classical nonlinear lattices associated with 
negative coupling constant $(c < 0)$.

\section{Iterated Maps}

Conversion between Hamiltonian forms and mappings has long been known 
to be a powerful mathematical tool for the theoretical and numerical 
analysis of dynamical systems.  Indeed, two-dimensional maps allow 
representation of a stationary configuration of a Hamiltonian of the form 
(\ref{adpot1}) by a trajectory of a dynamical system.  Therefore, the 
mathematical situation is {\it identical} to that of temporal evolution, 
although the static problem has been formulated in terms of spatial 
arrangements.

For a very small number of lattice sites we could attempt to solve 
the set of equations produced for each system.  To classify the 
type of solution obtained we analyse the wave vector structure of 
this solution.  Alternatively, for a larger number of sites,  we 
can follow the methods used in the studies of 
nonlinear dynamical systems: we can iterate the NSE (\ref{dns4}), 
after first representing it in the form of a $2D$ map.
That is, we can apply adapted methods from the analysis of chaotic 
dynamical systems and explore the nature of possible 
solutions to the NSE.  To do this we must first find suitable orbits.

The discretised nonlinear Schr\"odinger equation may be cast to form 
a first order $2-D$ iterative map in discrete time by introducing 
the auxiliary variable $Z_{i+1} 
= \psi_{i+1} - \psi_{i}$ (see for comparison \cite{KusK, Bak, Aubry}).  
This gives

\begin{equation}  
  \left( \begin{array}{c} 
  Z_{i+1}е \\
  \psi_{i+1}е 
  \end{array} \right) 
  = 
  \left( \begin{array}{c} 
  Z_{i} - E\psi_{i} - C\psi_{i}^3 \\
  \psi_{i} + Z_{i+1}
  \end{array} \right)  \qquad i = 0,1,\ldots
\label{2D-map}  
\end{equation}

\noindent which can also be considered as a three-term recursion.  The 
total spectrum of solutions depending on the initial conditions 
$\psi_{0}е$ and $Z_{0}е$ cannot be obtained in closed form.  In the 
early eighties it was common to generate possible 
solutions of (\ref{2D-map}) by fixing the displacements of two 
neighbouring particles and then iterating the map, however, the physical 
stability of the system under small perturbations was often not considered.  
In fact, the map technique provides a variety of stationary 
solutions of equation (\ref{dns4}).  However, the accessibility to 
{\it all} possible solutions is obviously highly limited by the parameter 
range where the map is mathematically stable.
Since the determinant of the Jacobian of the map equation 
(\ref{2D-map}) is strictly one, the map is area-preserving and the 
stability of a cycle ${\bf \psi} = \{\psi_{1}е, \psi_{2}е,\ldots,\psi_{N}е\}$ 
is guaranteed only if the eigenvalues of the Jacobian product along the 
trajectory are on the unit circle, ie the trace of the Jacobian of the 
$N-th$ iterate satisfies

\begin{equation}
  | {\rm Tr} \prod^{N}_{i=1}ее
  \left( \begin{array}{cc} 
  2 - E - 3 c \psi_{i}е^{2}е& -1 \\
  1 & 0 
  \end{array} \right) 
  | < 2 \label{2dmap2}
\end{equation}

When equation (\ref{2dmap2}) is not satisfied the trajectory 
$\{\psi_{i}е\}$ is unstable with respect to the initial conditions.
Note that the stability is only local in the sense that for initial 
conditions sufficiently close to the fixed point of the $N-th$ iterate, 
the corresponding phase portrait $\{(\psi_{i}е, Z_{i}е)\}$ describes the 
usual elliptic Kolmogorov-Arnold-Moser (KAM) curves.  Consequently, 
$N-$periodic solutions, defined by a fixed point of the $N-th$ 
iterate, depend singularly on the initial conditions and can only be 
obtained by providing the exact initial conditions.

As an example, consider the fixed point of the evolution of 
(\ref{2D-map}): inserting $(\psi^*,Z^*) = (\pm \sqrt {- {E \over c}},0)$ 
into (\ref{2dmap2}) we have

\begin{equation}
  | 2 + 2E | < 2 \label{2dmap3}
\end{equation}

\noindent The fixed point is only locally stable if $-2 < E < 0$.

Note that the stability behaviour does not depend on the 
parameter $c$; $c$ can be rescaled.  The 
normalisation condition, (\ref{nc1}), 
can be satisfied by a suitable rescaling method with the 
aid of the free parameter $c$.  If one rescales the wave function 
$\psi_{i}$ by $\beta$ and the coupling constant $c$ by $1/\beta^2$ the 
nonlinear Schr\"odinger  equation remains unchanged.  In other words, 
the value $\psi_{i}е$ and the value $c$ are related by the similarity 
transformation: $\psi_{i} \rightarrow \beta\psi_{i}$ and $c\rightarrow 
c/\beta^2$ \cite{Kus-Rash}.  Therefore we use the scaling parameter 
$\beta$ to satisfy the normalisation condition.  The  physical value 
of the (electron-phonon) coupling constant is then $C=c \sum_{i} \psi_{i}^2$.  

The next step is to fix the values $E$ and $C$ and then 
iterate the map.  Provided that for arbitrary values of $c$ a 
non-divergent solution $\psi_{i}$ exists, it will always correspond to 
a physical constant $C=c \sum_{i} \psi_{i}^2$ and the physical rescaled 
wave function is  $\Psi_i=\psi_{i}/ \sqrt{\sum_{i} \psi_{i}^2}$.

It is possible that for some value of $E$ and a 
certain lattice size that the iteration procedure is divergent.  In 
the classical theory of chaos this is not allowed because the phase 
space must necessarily be bounded.  However, in the quantum problem 
this is not important since the system has a finite size.  For any 
finite system we simply exclude divergent trajectories from our 
consideration because this divergence means that at such values of 
the parameters a solution does not exist.

Numerical experiments investigating different trajectories 
of the 2-D map have been performed for different values of the 
parameters.  We find that there are 
essentially only three types of maps produced by iteration.  
There can be {\it regular} maps where only a small number of points 
in the phase space are visited.  There are also {\it irregular commensurate} 
maps in the form of closed loops.  These loops consist of a number 
of points being visited.  As the commensurability 
decreases the points visited in the phase space become more and more dense 
and an elliptic orbit is mapped.  Finally, we have {\it irregular incommensurate} 
maps where the previous closed loops are dispersed (to the stochastic) 
and a large number of points in the phase space are visited.

These simulations indicate that in 
this system there are three qualitatively different types of 
trajectories which depend on the values of the parameters: 
1) periodic, 2) quasiperiodic and 3) chaotic.  It is therefore 
possible to speculate that 
the regular commensurate solutions may be transformed into 
the quasiperiodic or chaotic type structures by a change of 
initial conditions or parameters.  
The type of structure which arises in place 
of the regular one depends on how strong the incommensurability is.  
For example consider a period 2 solution.  If there is an even number 
number of sites then this solution is commensurate with the lattice.
However, if there is an odd number of sites then there is one site too 
many for the period 2 solution to exist.  In this case the 
incommensurability which is caused by the appearance of this site 
is distributed throughout the lattice but is mostly seen 
near the extra site.  If 
there is a large number of lattice sites then the incommensurability 
caused by the extra site is weak and so the possible solution arising 
(instead of the regular one) 
will be a quasiperiodic one.  The period 2 
structure will be slightly deformed. 
This quasiperiodic solution 
(with respect to period 2)  will at the same time also be 
a regular period $N$ solution if we are considering 
PBC. This is true for any other quasiperiodic or 
chaotic solution of NSE.

The incommensurability may be stronger in other cases.
For example, consider a period $m$ structure on an $N$ site 
lattice $(m < N)$ where $N$ is prime.  If $k$ such period $m$ structures 
exist on this lattice then the number of sites can be decomposed as 
$N = k m + k_{1}$ 
with $k_{1} \simeq {m \over 2}$ is a number of the order of $k$.  
Then the incommensurability arising from
the extra $k_1$ sites will deform the period $m$ solution. 
If this incommensurability is sufficiently strong then 
we would expect to have chaotic solutions arise 
instead of the regular period $m$ solutions.  

Below we show that these qualitative arguments are indeed valid.  
To prove that the proposed classification
does exist let us try to find all possible solutions
of the NSE for a finite system directly applying
analytic and numeric methods.

\section{Exact Solutions in the limit $c \rightarrow \infty$}

The above NSE (\ref{dns4}) has exact 
solutions in the limit $c \rightarrow \infty$ for a $N$-site lattice,
which have been published separately \cite{KusD}.  
The associated energy eigenvalue of these exact solutions takes the form 
(for $c > 0$)

\begin{equation}
    E = {2m + 4l - c \over n}    \label{eieq}
\end{equation}

\noindent where the 
eigenvalue $E$ corresponds to a wave function 
localized with near equal probability 
on $n$ sites 
which are separated into $m$ groups (spots).  Between the spots
the wave function is vanishing while 
inside these spots the wave function changes sign a total 
of $l$ times, although its amplitude is (nearly) the same.  
Note that in each localizing quantum well the wave function 
is localized with equal probability, $\psi_{k}^{2} = 1/n$ for all 
numbers $k$ associated with localizing quantum wells. This probability does
not
depend on the number of spots, $m$, 
the localization area
is separated into.    The number $n$ may take any integer 
value $n = 1,2,3,\ldots,N$.  
The number $m$ may take any integer value satisfying 
$m \leq {n}$.  
Similarly, the number $l$ satisfies $l \leq n$.

Eq. (\ref{eieq}), 
obtained in the limit $c \rightarrow \infty $, has been compared 
with the exact and numerical 
solutions for systems consisting of  small number of sites.
In all these cases for nearly all values of $c$ (except small regions
of critical values where the self-trapped solutions originate) 
there is perfect agreement
with the derived formula (\ref{eieq}).
 However, in contrast with this perfect agreement between
 eigenvalues, a decrease in the value of $c$ leads to a noticeable
 deviation in the  
wave functions (eigenvectors)  from those  
 obtained in the limit $c\rightarrow\infty$.
The first order corrections to the presented eigenvectors obtained with
the use of perturbation theory is of the order
of $\sqrt{n}/c$ ie $O(1/c)$. When the coupling constant $c$ is not very large
 the wave functions of some states
will have interesting incommensurate and
chaotic structures. 
Thus, from comparison with numerical results
and  with perturbation theory we conclude, 
that even though the spectrum of the NSE for a system 
with a finite, arbitrary number, $N$, of sites 
is well described by equation (\ref{eieq}) for $c \gg 1$, 
the shape of the appropriate 
wave function for smaller values of $c$ may have 
only qualitative features of the wave function 
obtained in the limit $c \rightarrow \infty$.  
As $c$ decreases the localization spots smear out.
Since the spectrum, eq.(\ref{eieq}), is associated with a local
localization pattern, it is universal and 
does not depend on the boundary conditions.

Although the structure 
of the wave function for $c \gg 1$ does not strongly agree with 
the structure of the wave function in the limit $c \rightarrow \infty$ 
we find that the wave function may be approximated reasonably well by 
an exponential function.  This is only valid if the peaks in the wave 
function structure are separated sufficiently so that there is little 
or no interaction between the tails.  For lattice sites sufficiently 
far away from the localised site we can assume that in (\ref{dns4}) the 
value of the wave function is small so that the cubic nonlinear 
term vanishes.  Then, in the continuum limit, we obtain 
a second order linear differential equation which may be expressed as 

\begin{equation}
  -{\partial^2 \psi \over {\partial x^2}} = E \psi(x)  \label{conteq1}
\end{equation}

(\ref{conteq1}) has the asymptotic solution 

\begin{equation}
  \psi = A \exp(- \sqrt{-E}еx)  \label{conteq2}
\end{equation}

\noindent where $A$ is some parameter which from the normalisation 
condition goes as $A \sim (-E)^{1/4}е$.   (\ref{conteq1}) may only 
be used to describe the behaviour of the wave function sufficiently far 
away from the local maxima of the wave function because near the 
maxima the cubic nonlinear term cannot be neglected.  Hence, away 
from the peak the behaviour of the wave function may be described as 
an exponential decay.

Each of the eigenvalues $E$ correspond to energies created due to 
a wave function localising in quantum 
wells.  Therefore, it is interesting to estimate
the Hamiltonian, $H$, which corresponds to the appropriate structures.  
With the use of the method described above  the following 
expression for the Hamiltonian, $H$, is obtained

\begin{equation}
     H = {c \over 2n}   \label{adpoteq}
\end{equation}

\section{Perturbation Theory}

If a small perturbation in the wave function (and energy eigenvalue) 
is considered, with $c \gg 1$, then a correction to the wave function 
is revealed.

Let 

\begin{equation}
    \psi_{i} = p_{i} + x_{i}  
\end{equation}

\noindent and 

\begin{equation}
    E = E_{0} + E_{1}
\end{equation}
    
\noindent where $p_{i}$ is the zero approximation wave function, 
$x_{i}$ is a small perturbation in the wave function, $E_{0}$ is 
the zero approximation energy eigenvalue and $E_{1}$ is a small 
perturbation in the energy eigenvalue.  Substituting this into 
(\ref{dns4}) and ignoring the smaller terms gives

\begin{equation}
    - p_{i-1} - x_{i-1} + 2 p_{i} + 2 x_{i} - p_{i+1} - x_{i+1} 
    - c p_{i}^{3} - 
    3 c p_{i}^{2} x_{i} \approx E_{0} p_{i} + E_{0} x_{i} + 
    E_{1} p_{i} + E_{1} x_{i}  \label{eq2}
\end{equation}

In matrix form the system of $N$ equations becomes 

\begin{equation}
  {\bf T X = F}  \label{veceq1}
\end{equation}  
  
\noindent where ${\bf F}$ and ${\bf X}$ are vectors with $N$ components 
($N$ is the number of lattice sites).  The 
components of ${\bf F}$ are $F_{i} = E_{0} p_{i} + 
c p_{i}^{3} + p_{i-1} - 2 p_{i} + p_{i+1}$, the components 
of ${\bf X}$ are $X_{i} = x_{i}$ ie ${\bf X}$ is the wave vector of first 
order corrections $(\sim 1/c)$ to the zero approximation wave function.  
Finally, (for 3 or more sites)  ${\bf T}$ is a tridiagonal $N$ x $N$ matrix 
with diagonal elements $T_{i,i} = 2 - E_{0} - 3 c p_{i}^{2}$, 
elements on either side of the diagonal $T_{i,i-1}, T_{i,i+1} = -1$ and 
top-right and bottom-left elements $T_{1,N}, T_{N,1} = -1$ for 
periodic boundary conditions (PBC).  The terms involving $E_{1}е$ have 
been ignored for the meanwhile because $|E_{0}е| \gg |E_{1}е|$.  $E_{0}е$, 
which is given by (\ref{eieq}), is in excellent agreement with the 
actual value of the energy eigenvalue for nearly all values of $c$.

Premultiplying equation (\ref{veceq1}) by ${\bf T^{-1}}$ gives the 
first order corrections to the wave vector:

\begin{equation}
  {\bf X = T^{-1} F}  \label{veceq2}
\end{equation}

Now that we have the first order perturbation corrections to the 
wave function, $x_{i}$, we can 
substitute these in (\ref{eq2}) to calculate the first order 
correction to the energy eigenvalue, $E$.  Assuming PBC and 
summing (\ref{eq2}) over all $i$ (neglecting the nonlinear 
terms in $x_{i}$ as $x_{i}^{2} \ll x_{i}$), we obtain the 
first order correction to the energy eigenvalue

\begin{equation}
   E_{1} = {- c \sum_{i} p_{i}^{3} - 3 c \sum_{i} p_{i}^{2} x_{i} 
          - E_{0} \sum_{i} p_{i} - E_{0} \sum_{i} x_{i} \over 
          \sum_{i} p_{i} + \sum_{i} x_{i}}  \label{enrgcorr1}
\end{equation}

To see how the perturbation corrections compare to actual solutions 
we compare these corrections with exact solutions of (\ref{dns4}).  For 
a 2-site system with PBC we know that the exact solutions correspond 
to the eigenvalues $E = - c/2, E = (8 - c)/2$ and $E = 2 - c$.  The 
corresponding eigenvectors are 

\noindent for $E= - c/2$ \hspace{10mm} 
\( \left[ \begin{array}{r}
 {1 \over \sqrt{2}} \\
 {1 \over \sqrt{2}}
\end{array} \right] \)

\vspace{5mm}

\noindent for $E= (8 - c)/2$ \hspace{10mm} 
\( \left[ \begin{array}{r}
  {1 \over \sqrt{2}} \\
 -{1 \over \sqrt{2}}
\end{array} \right] \)

\vspace{5mm}

\noindent for $E=2 - c$ \hspace{10mm} 
\( \left[ \begin{array}{r}
 {\sqrt{1 \pm \alpha} \over \sqrt{2}} \\
 {\sqrt{1 \mp \alpha} \over \sqrt{2}}
\end{array} \right] \)

\vspace{5mm}

\noindent where $\alpha = \sqrt{1 - {16 \over {c^2}}}$ and lies in the 
range  $0 \leq \alpha \leq 1$.  This solution only exists for 
$c > 4$.  The above three eigenvalues (obtained analytically) are 
precisely the same as those given by (\ref{eieq}) if the appropriate 
wave function structures are substituted.  Note that for 
$E =- c/2$ and $E = (8 - c)/2$ we have no spots ie $m=0$.  There are no 
sites where the wave function is zero.  If Open Boundary Conditions 
(OBC) were considered (instead of PBC) then, due to the nature of the 
boundary conditions, $m \neq 0$

For the solutions which correspond to the eigenenergies $E =- c/2$ and 
$E = (8 - c)/2$ we know that there are no corrections to the 
wave function and energy eigenvalue.  Thus, we expect that $x_{i} = 0$ 
and $E_{1} = 0$ for both of these solutions.  Substituting for $E_{0}$ 
and $p_{i}$ in (\ref{eq2}), (\ref{enrgcorr1}) we find that this is 
indeed the case.  

Following the same procedure for the symmetry breaking solution gives 
no first order correction to the localised component of the wave 
function (but does give a correction of $-2/c^{2}е$ 
for $c \gg 1$) and a first order correction of 
$2/c$ to the vanishing component.  To compare these results with the 
exact solutions we must expand the surd (in inverse powers of $c$) 
for both the localised and vanishing components of the wave function.  
Doing so gives 

\begin{eqnarray}
    \sqrt{1 + \sqrt{1 - {16 \over {c^2}}}} \over \sqrt{2} &\approx& 
    1 - {2 \over c^{2}} - {10 \over c^{4}} - {84 \over c^{6}} - \ldots  
    \label{2sitea} \\ 
    \sqrt{1 - \sqrt{1 - {16 \over {c^2}}}} \over \sqrt{2} &\approx&
     0 + {2 \over c} + {4 \over c^{3}} + {28 \over c^{5}} + {264 \over c^{7}} 
     + \ldots  \label{2siteb}
\end{eqnarray}

\noindent which agrees completely with the results from our 
perturbation analysis.  Note that in the limit $c \rightarrow \infty$ 
the corrections are all zero and the zero approximation 
solutions are obtained.  Although there is excellent 
agreement in the wave functions, the same is not true regarding the 
energy eigenvalue.  We know that the eigenvalue $E = 2 - c$ is exact 
and consequently there should be no corrections to $E_{0}$.  
However, we obtain the correction $E_{1} \approx - 4/c$.  This is 
because we did not take into consideration  all orders of correction.  
We have only used first order corrections and neglected lower 
order corrections.  Clearly, lower order corrections
do exist and these too must be considered for complete 
agreement.  Thus, the correction in the energy eigenvalue, $E_{1}$, 
is zero to an order of $c^{0} \sim 1$.

Once we have obtained the first order corrections to the wave function 
and eigenenergy we can then investigate higher order corrections.  From 
the 2-site PBC example discussed above we expect that the next correction 
to the wave function will, in general, be of order $\sim 1 / c^{2}$. 

\section{Iteration Procedure}

We can 
modify the above procedure by finding the first order correction to the 
zero approximation wave function obtained in the limit $c \rightarrow \infty$ 
as detailed above.  We then find the next order correction 
to this new wave function.  That is we repeat the above procedure and 
determine the next order correction to the `corrected' wave function.  
This iterative procedure (which is reminiscent of the Newton-Raphson 
procedure) is repeated until 
the correction to the wave function becomes negligible.   This means 
that the value of the wave function converges to some limit.  This limit 
is an exact, numerical solution of the discretised NSE (\ref{dns4}).

This exact solution is obtained by following an analogous procedure to 
that used to obtain (\ref{veceq2}).  If $\psi_{i}(m)$ is our 
wave function iterated $m$ times then we again have 

\begin{equation}
  {\bf T X = F}  \label{veceq3}
\end{equation}  
  
\noindent where ${\bf F}$ and ${\bf X}$ are again vectors with $N$ components 
associated with the number of lattice sites.  The 
components of ${\bf F}$ are now $F_{i} = E_{0} \psi_{i}(m) + 
c \psi_{i}^{3}(m) + \psi_{i-1}(m) - 2 \psi_{i}(m) + \psi_{i+1}(m)$ 
and the components 
of ${\bf X}$ are $X_{i} = x_{i}$ the $(m+1)$th iterative 
corrections to the wave function iterated to $m^{th}$ order.  
Finally, ${\bf T}$ is again a tridiagonal $N$ x $N$ matrix with diagonal 
elements $T_{i,i} = 2 - E_{0} - 3 c \psi_{i}^{2}(m)$.

Premultiplying equation 
(\ref{veceq3}) by ${\bf T^{-1}}$ gives the $(m+1)$th iteration corrections 
to the wave vector:

\begin{equation}
  {\bf X = T^{-1} F}  \label{veceq4}
\end{equation}

Alternatively, we may write the linear approximation to the wave function 
as

\begin{equation}
  \psi(m+1) = \psi(m) + \delta \psi(m)
\end{equation}

\noindent and analogously 

\begin{equation}
  E(m+1) = E(m) + \delta E(m)
\end{equation}

\noindent so that the counterpart of (\ref{eq2}) after $m$ iterations 
may be written in the form

\begin{equation}
 \hat \nabla H ({\bf \Psi} (m+1)) - \hat \nabla H ({\bf \Psi} (m)) 
 \approx \hat \nabla^{2}е H ({\bf \Psi} (m)) ({\bf \Psi} (m+1) - {\bf \Psi}(m)) 
\end{equation}

Then (\ref{veceq3}) can be expressed in the form

\begin{equation}
  \hat \nabla^{2}е H ( {\bf \Psi} (m)) \delta {\bf \Psi(m)} = 
  \hat \nabla H ({\bf \Psi} (m)) 
\end{equation}

\noindent with ${\bf F} = \hat \nabla H ({\bf \Psi} (m))$, 
${\bf T} = \hat \nabla^{2}е H ({\bf \Psi} (m))$ and ${\bf X} = \delta 
{\bf \Psi}(m)$ so that 

\begin{equation}
   {\bf \Psi} (m+1) = {\bf \Psi(m)} - \left[ \hat \nabla^{2}е H ({\bf \Psi} 
   (m)) \right]^{-1}е \hat \nabla H ({\bf \Psi} (m)) 
\end{equation}

This process is, however, not always convergent.  The wave vector may 
change structure when other eigenenergies have a similar value 
for a particular $c$.  For example, for a 3 site system with PBC 
two of the eigenvalues are $-c/3$ and $(6-c)/2$.  Both of these 
have the same value for $c=18$ but they have very different structures 
of the wave function.  It is possible for 
the initial wave vector structure corresponding to one energy eigenvalue 
to change during the iteration process to another structure which corresponds 
to a different energy eigenvalue.  However, this new structure does also 
exist and is also a solution of the NSE (\ref{dns4}).

Analogous to (\ref{enrgcorr1}), the energy eigenvalue (for PBC) is given 
by 

\begin{equation}
  E(m) = \frac {-c \sum_{i} \psi_{i}^{3} (m)} {\sum_{i} \psi_{i} (m)}
  \label{enrgycorrm}
\end{equation}

\noindent and the value of $E(m)$ should converge towards $E_{0}$ 
for most values of $c \gg 1$ except for small regions of $c$ where the 
localised states originate.  Even in these sensitive regions the value 
of $E(m)$ should be in fairly close agreement with $E_{0}$.  If there is a 
drastic change in the value of $E(m)$ then this indicates that the wave 
function structure has changed and the iteration process will converge to 
some other limit.

Let us apply this method to the above NSE for finite $c$ taking 
the zero approximation solutions, which are exact in the limit 
$c \rightarrow \infty$, as the starting condition.  The results obtained 
by the Iteration Procedure can be 
compared with the results obtained by the use of $2D$ iterated maps. 
For such a comparison, however, we have to introduce the new {\it phase 
space function} 
$Z_{i} = \psi_{i+1} - \psi_{i}$ for the Iteration Procedure and 
plot $Z_{i}(\psi_{i})$ against $\psi_{i}$. This then allows us 
to compare the wave function structures obtained with the use of the 
Iteration Procedure and the trajectories of the 2D iterated maps 
discussed in Section 3.  We find that the results from the two different 
methods are consistent.  That is the pictures representing the phase 
space of our quantum system correspond to different periodic or chaotic 
trajectories of the Poincar\'e Map.

\section{Results}

In the previous section we have reformulated our problem of the 
solution of a discrete nonlinear set of equations, (\ref{dns4}), 
to an iterative problem.  This gives 
possible criteria for the classification of the solutions obtained.
Following the Iteration Procedure we find that there exist different types of 
structures of the wave function.  We find that at some fixed initial 
conditions for $E>0$ there are the usual plane wave Bloch solutions 
(analogous to the motion of a free electron).  The corresponding phase 
portrait has the form of a closed (elliptic) curve.  However, nonlinear 
localized solutions arise for $E<0$.  
Our results indicate that there are three qualitatively 
different types of structure of the wave vector for negative values of $E$: 
periodic, quasiperiodic and (deterministic) chaotic.  The same structures 
can also be obtained by numerically solving the NSE for a finite size system.  
Of course for finite systems quasiperiodic and chaotic structures are not 
well defined because these are, strictly speaking, well defined only for 
systems of infinite size.  However, we may still indicate 
analogous features, for example, with the aid of phase portraits 
adopted for finite size systems which are possibly 
equivalent to commensurate, regular incommensurate and irregular 
incommensurate structures of the wave function on the lattice.

\subsection{Periodic Trajectories}

Consider the initial structure where the normalised wave function is 
located on only 4 equidistant sites, ie there are only 4 sites where 
the wave function is not zero.  These 4 sites are 
equally spaced apart on a lattice consisting of a total of 32 sites with 
PBC.  Each of the localised sites is separated by 7 sites where the 
wave function is zero.  The corresponding 
energy eigenvalue is given by $E = (8 - c)/4$.  
Taking the coupling 
constant to be $c = 10$ and applying the Iteration Procedure to this 
initial structure of the normalised 
wave function we obtain the solution presented in 
Figure 1a.  This shows a periodic behaviour of the wave function 
through the lattice.  The corresponding phase portrait of the quantum 
state, Figure 1b, indicates that this initial structure gives 
rise to a regular commensurate period 8 structure.  The 
phase portrait consists of only 8 points as the wave function 
oscillates regularly.  This solution represents a generalisation of 
the period two solution for our two site system which we obtained 
analytically.  Here the 
wave function will occupy each 8th site with a greater probability than 
the other 7 sites.  Since the lattice deformation is proportional to 
$\psi_{i}^2$, each 8th site  will be deformed by a greater amount.  
Such a regular deformation will give rise to a superlattice of 
deformations or a superlattice of solitons.  As $c$ is increased 
the oscillations in the value of the wave function will grow.  
The width of each peak decreases and the height increases.  
The wave function is a maximum at every 8th site.  In the limit 
$c \rightarrow \infty$ we will obtain our initial structure of the 
wave function being completely localised on each 8th site and zero 
everywhere else.

\subsection{Quasiperiodic Trajectories}

If the initial structure is altered slightly so that the localised 
sites are not equidistant then we see a deviation from the regular 
periodic behaviour of the wave function.  Figure 2a depicts the 
wave function, obtained via the Iteration Procedure, 
which corresponds to an initial condition for the Iteration Procedure  
consisting of 4 localised sites on a 32 site lattice with PBC.  
However, this time the localised sites are separated by either 5 or 
9 lattice sites.  The energy eigenvalue is still given by $E = (8 - c)/4$.  
For $c = 12$ we see some competition develop between 
the different localised sites.  They begin to interfere with each other 
and give an irregular structure which has competing periods but is, 
nevertheless, commensurate with the lattice.  
These periods deviate slightly from one another.  This is illustrated in 
Figure 2b where we see a dispersion of the points in the phase 
space.  The previous regular period 8 structure no 
longer exists on the lattice.  Two different period 8 structures arise 
which correspond to two different sets of points being visited in the 
phase space.   The deviation in the period 8 structures indicate the 
origins of quasiperiodicty or even chaoticity within the system due to 
a strong interaction of different spots or different solitons. 
Specifically, it is easy to 
imagine a larger lattice with a greater number of competing 
structures.  The phase space of the quantum system will then {\it become 
more dense and take the form of a closed loop which would be indicative 
of a quasiperiodic solution.  If this closed loop is dispersed then this 
would indicate a stochastic structure which is reminiscent of classical 
chaotic motion.}

If the value of the energy eigenvalue becomes less negative, ie 
closer to zero but still negative (for example $c = 11$) then the 
interaction between the spots intensifies.  Similarly, if the value of 
$c$ is increased then the interaction between the spots is reduced.  
In the limit 
$c \rightarrow \infty$ we will obtain our initial structure of the 
wave function being completely localised on 4 sites and zero 
everywhere else.  Note that in this limit 
the corresponding phase space will consist 
of just 3 points: one at the origin $(\psi_{i}, Z_{i}) = (0, 0)$, 
one at $(\psi_{i}, Z_{i}) = (0, 1/ \sqrt{n})$ and one at $(\psi_{i}, 
Z_{i}) = (1/ \sqrt{n}, -1/ \sqrt{n})$.

Figure 3a corresponds to the final state of a wave function which 
had the initial condition for the Iteration Procedure of the wave 
function being localised on 10 
lattice sites which are all grouped together in one spot on a 
100 site lattice with PBC.  The energy eigenvalue is given by 
$E = (2 - c)/10$.  For $c = 4$ the interaction between the 
localised sites gives rise to the wave function being localised 
in two discrete spots.  Note that the value of the wave function is 
always positive.  The phase space, Figure 3b, indicates a possible 
quasiperiodic (or regular incommensurate) structure.  The phase space of 
this system lies on a closed 
loop.  The points which belong to this phase space appear to be paired.  
This could be an indication of small deviations in the structures of 
the wave function of the two spots.  However there is a dispersion 
of the closed loop trajectory at the origin.  This dispersion indicates 
a transition from a quasiperiodic structure to a weakly chaotic regime.  
The dispersion is due to a commensurate wave function 
structure beginning to transform into an incommensurate structure.

Figure 4a shows the final wave function structure, obtained via the 
Iteration Procedure, which had the initial 
condition of the wave function being localised on 12 lattice sites in 
12 different spots on a 100 site lattice with PBC.  The number of empty 
lattice sites separating the localised sites varies cyclically.  There were 
6, 7 and 9 empty lattice sites between different localised sites.  This 
pattern was repeated systematically every 25 sites.  The 
energy eigenvalue is given by $E = (24 - c)/12$.  For $c = 29$ we obtain 
a wave function pattern, Figure 4a, which at first seems to have no 
order.  Figure 4b shows the underlying pattern inherent in the wave 
function structure.  The phase portrait of this quantum system, Figure 4c, 
indicates a period 25 structure.  This is due to the structure 
being repeated every 25 lattice sites.  From Figure 4b we see that 
there are actually three peaks of the wave function which occur over 
25 lattice sites.  That is 
the loop plotted in Figure 4c must actually include 3 trajectories which 
span the phase space of the phase portrait.  Thus, in Figure 4c we see 
something akin to a regular incommensurate structure of the wave function.

\subsection{Chaotic Trajectories}

Figure 5a shows the behaviour of the wave function on the lattice sites 
as $c$ is increased to $c = 31$ for the previous initial condition.  
The energy eigenvalue $E = (24 - c)/12$ becomes more negative.  The 
wave function structure has split into three regions; 
top, middle and bottom.  Figure 5b shows that some type of symmetry has 
been imposed on the wave function.  The structure is repeated every 25 
sites as expected.  In each set of 25 lattice sites the wave function 
has three local maxima.  Two of these maxima consist of just one lattice 
site but one local maxima consists of two lattice sites.  This is due to 
the strong interaction of the spots.
The corresponding phase space, Figure 5c, is in the form of another dispersed 
closed loop.

As $c$ is further increased to $c = 32$ Figure 6a shows the wave 
function structure which has now further separated into very distinct 
groups.  From figure 6b one can see that now the interactions between 
the spots has decreased.  The peaks are narrower and higher.  Figure 6c 
represents the phase space of this structure 
which is comprised of sets of three points.  
This chaotic structure is due to the interaction of different localisation 
spots.  The appearance of the stochastic layer demonstrates 
a chaotic behaviour and indicates the effects of irregular incommensurability 
of the soliton lattice with the original  atomic lattice.  Upon 
further inspection of Figure 6a, which shows 
the value of the wave function through out the lattice, we see that 
the different horizontal regions in Figure 6a actually 
correspond to the different points in the orbit of the phase portrait, Figure 
6c.  This orbit has 8 sets of three points and on closer inspection of 
Figure 6a we see that Figure 6a is actually organised into 8 horizontal 
sets; one at the top, two in the middle and five at the bottom.

Figure 7a shows the behaviour of the wave function on the lattice sites 
as $c$ is increased to $c = 36$.  The energy eigenvalue $E = (24 - c)/12$ 
becomes more negative and the interaction between the spots decreases.  
The wave function is again divided into three distinct regions 
at the top, middle and bottom.  
These distinct regions correspond to the dispersed points in 
Figure 7b.  The dispersion of these points has now 
reduced and they are beginning to converge 
towards different limits.  This indicates that the incommensurate structure 
is becoming a commensurate structure.  The 
wave function pattern in Figure 7a is separated into different sets 
which correspond to the different convergence limits in Figure 7b.  
In the limit $c \rightarrow \infty$ the phase space will converge 
to the points $(\psi_{i}, Z_{i}) = (0, 0)$, 
$(\psi_{i}, Z_{i}) = (0, 1/ \sqrt{n})$ and $(\psi_{i}, 
Z_{i}) = (1/ \sqrt{n}, -1/ \sqrt{n})$.

As $c$ is further increased the wave function structure now begins to 
resemble the initial condition where the wave function was localised on 
12 sites in 12 spots.  Note that this initial structure can only exist 
in the limit 
$c \rightarrow \infty$.  Figure 8a shows the wave function of the lattice 
sites for $c = 84$.  The energy eigenvalue $E = (24 - c)/12$ becomes even 
more negative.  The value of the wave function now has three 
specific domains; at the top, near the bottom and at the bottom.  These 
domains correspond to the different states of the system as depicted in 
Figure 8b.  The wave function structure now indicates regular 
commensurability.  In actual fact, careful study of figure 8b indicates a 
period 8 structure; there are three points closely located at the origin.  
If the resolution of Figure 8b was better it would 
actually indicate a period 8 structure.  Note that the sets of three points 
which caused the dispersion of the orbit (see Figure 6c) have converged 
as $c$ was increased and now there is only one discrete point where 
previously there were three.  This dispersion of the points for 
relatively small $c$ could possibly be due to the slightly different 
corrections of each order to the wave function.  This is indicated 
by the 18 possible first order corrections (see appendix A1) depending 
on the structure of the wave function.  The first order correction 
for different sites are not all the same.  These differences can cause 
a variation in the pattern of the wave function which would result in 
a dispersion of the orbit of a trajectory in the phase space.  
It is easy to extrapolate this result for different orders of correction.  
For each order of correction to the wave function there are an increasing 
number of possible corrections.  More and more different corrections appear 
which decrease for large $c$ and in the limit $c \rightarrow \infty$ these 
corrections vanish.  This then could be the origins of the irregular 
incommensurability present in our quantum system.

\section{Summary}

Thus, we obtain that in the limit $c \rightarrow \infty$ there arises a 
degenerate set of solutions associated with different localisation 
patterns of the wave function which consist 
of empty lattice sites (where the wave function is vanishing) and lattice 
sites where the wave function is localised.  These localisation patterns, 
which can be viewed as soliton type structures, have 
many different configurations which correspond to the same eigenvalue of 
the NSE.  However, when the value of $c$ is not infinite this degeneracy 
is broken.  Different localisation spots within the pattern start to 
interfere with each other and modify the behaviour of the wave function.  
The interaction between the solitons is 
determined by the reciprocal of the modulus of an eigenvalue of the NSE, 
$1/ \sqrt{|E|}$, which governs the radius of the soliton peaks.   

The interaction between the solitons depends essentially on the radius of 
the solitons and their separation.  
If the radius of the soliton peaks is much smaller than the separation 
then we have regular, commensurate behaviour of the wave function 
on the lattice.  The interaction between the solitons is very weak.  
As the radius of the solitons increases 
($1/ \sqrt{|E|}$ increases) and the separation remains constant the 
interaction between the solitons increases and the behaviour of the 
wave function on the lattice becomes less regular.  Finally, when the 
radius of the soliton peaks is about the same as the separation of the solitons 
we have strong interactions between the solitons.   This results in 
strong incommensurability of the behaviour of the wave function on 
the lattice.

We find that the interference between the spots can give rise to 
three qualitatively different structures: periodic, quasiperiodic and
chaotic.  To indicate such structures in the quantum state the methods 
used in the studies of classical chaos were applied to the quantum 
system.  A \emph{Poincar\'e map for the quantum state} was built up.  
This consists of the amplitudes of the wave functions and of residues 
of these amplitudes associated with neighbouring sites.  That is it 
is a projection of the Hilbert  `phase space of our quantum system' 
in the plane, that is, the set $\{Z_{i}, \psi_{i}\}$.  

In the regular periodic and quasiperiodic solutions the 
wave function amplitudes replicate with some period equal to some 
integer number of lattice constants or creates aperiodic structures, 
respectively.  
However, there is also the appearance of structures analogous to the 
those arising in classical chaos.  
The destruction of the periodic and quasiperiodic 
orbits, which has been ascribed to the creation of chaotic structure 
in the wave function, is also exhibited by the system.  That is the 
creation of a structure which has no definite period has been found.

\vfill

\eject

{\bf Figure Captions}

\

Figure 1 The final (converged) behaviour of the wave function 
on the lattice (a)
and the corresponding phase portrait (b) obtained by Iteration 
Procedure for a 32 site lattice with Periodic Boundary Conditions 
(PBC).  The initial conditions were $c = 10$ and $E = -0.5$ with 
the initial, normalised wave function value of $\psi = 1/2$ at 
only 4 equidistant sites.  Between these `localised' sites the value 
of the wave function was zero.  Figure 1a shows a regular (periodic) 
behaviour of the wave function.  The pattern in Figure 1b indicates 
that only 8 points are visited all of which lie on a closed loop.  
Therefore, this structure of the wave function corresponds to a 
regular class of behaviour.

\

Figure 2 The final behaviour of the wave function on the lattice (a)
and the corresponding phase portrait (b) obtained via the Iteration 
Procedure for a 32 site lattice with PBC.  The initial conditions 
were $c = 12$ and $E = -1$ with the initial, normalised wave 
function value of $\psi = 1/2$ at only 4 sites.  These 4 sites 
were not distributed uniformly throughout the lattice but had either 
5 or 9 sites where $\psi = 0$ giving an `irregular' structure.
Figure 2a shows the irregular behaviour of the wave function.  
Figure 2b shows the 16 points which are visited in the phase 
space.  The points no longer lie on a closed loop and so this 
arrangement of points is associated with an irregular class of 
behaviour of the wave function.

\

Figure 3 The final behaviour of the wave function on the lattice (a)
and the corresponding phase portrait (b) obtained via the Iteration 
Procedure for a 100 site lattice with PBC.  The initial conditions 
were $c = 4$ and $E = -0.2$ with the initial, normalised wave 
function value of $\psi = 1/\sqrt{10}$ at 10 sites.  These 10 sites 
were all gathered together in one spot (group of localising sites).  
Figure 3a shows the behaviour 
of the wave function after it has converged.  Figure 3b shows the 
trajectory of this wave function.  Again a closed loop is 
traced out by the points visited in the phase space, however, this 
time there is some dispersion of points at the origin.  The loop 
appears to consist of pairs of points; each point in a pair 
corresponding to the change in the wave function at equivalent 
locations in either of the two separate peaks observed in Figure 3a.  
This arrangement of points 
in Figure 3b indicates some type of irregular structure of the 
wave function which is somehow commensurate with the lattice.

\

Figure 4 The final behaviour of the wave function on the lattice (a)
and the corresponding phase portrait (b) obtained via the Iteration 
Procedure for a 100 site lattice with PBC.  The initial conditions 
were $c = 29$ and $E = -0.416^{.}е$ with the initial, normalised wave 
function value of $\psi = 1/\sqrt{12}$ at 12 sites.  These 12 sites 
were separated into 12 spots which were further separated by either 
6, 7 or 9 `empty' sites.  This structure was repeated every 25 sites.  
Figure 4a shows the apparantly random behaviour of the wave function 
after it has converged.  Figure 4b shows the underlying pattern 
manifest in this structure.  The pattern is repeated every 25 lattice 
sites as expected.  Figure 4c shows the 
trajectory of this wave function.  Once again a closed loop is 
traced out by the 25 points visited in the phase space.  However, 
this loop corresponds to three distinct orbits; one for each of the 
three peaks in the wave function every 25 lattice sites.

\

Figure 5 As Figure 4 but with $c = 31$ and so $E = -0.583^{.}е$.  
Figure 5a shows the apparantly random behaviour of the wave function 
after it has converged.  Figure 5b shows the underlying pattern 
manifest in this structure which repeats itself every 25 lattice sites.  
Figure 5c shows that the 25 points visited in the phase space no longer 
lie on a closed loop.  However, most of these points seem to belong to 
a set of three points.  This could again (see Figure 3) correspond to 
the change in the wave function at equivalent locations in the separate 
peaks observed in Figure 5b.  This arrangement of points possibly indicates 
some type of irregular structure of the wave function which is 
incommensurate with the lattice.

\

Figure 6 As Figures 4 \& 5 but with $c = 32$ and so $E = -0.6^{.}е$.  
Figures 6a and 6b show the behaviour of the wave function 
after it has converged and the underlying pattern 
manifest in this structure, respectively.  
Figure 6c shows the 25 points visited in the phase space.  It can 
now be clearly seen that most of these points occur in sets of three.  
As the peaks in the wave function get narrower and higher (Figure 6b) 
the dispersion of these points decreases.  The behaviour of the wave 
function in Figure 6a seems to be divided into zones at the top, middle 
and bottom.  These zones correspond to the sets of different points 
in Figure 6c.

\

Figure 7 As Figures 4, 5 \& 6 but with $c = 36$ and so $E = -1$.  
Figure 7a shows the behaviour of the wave function 
after it has converged.  
Figure 7b shows the 25 points visited in the phase space.  The 
zones in Figure 7a are much more distinct and further apart as 
$c$ increases.

\

Figure 8 As Figures 4, 5, 6 \& 7 but with $c = 84$ and so $E = -5$.  
Figure 8a shows the behaviour of the wave function after it has 
converged.  Figure 8b shows the points visited in the phase space.

{\bf Appendix A1}

As the first order corrections to the wave function 
depends mainly on the value of the zero approximation wave function 
at only three neighbouring sites it is simple to obtain algebraic 
expressions for all first order corrections to the wave function.  
Note that, although the first order correction to the wave function 
at a particular site 
does have some dependence on the first order corrections to the 
wave function of its neighbouring sites, because $x_{i-1} \ll p_{i-1}$ and 
$x_{i+1} \ll p_{i+1}$ this dependence is weak and can, often, 
be ignored.  
In (\ref{eq2}), $p_{i-1}, p_{i}, p_{i+1}$ may be either 0, $1/\sqrt{n}$ or 
$-1/\sqrt{n}$ (assuming $x_{i-1}, x_{i+1} \sim 0$).

For $p_{i} = 0$:

$x_{i} = 0$ if both $p_{i-1}$ and $p_{i+1}$ are $0$.

$x_{i} = {\sqrt{n} \over (c - 2m - 4l)}$ if either $p_{i-1}$ is 
$1/\sqrt{n}$ and $p_{i+1}$ is $0$ or vice versa.

$x_{i} = {-\sqrt{n} \over (c - 2m - 4l)}$ if either $p_{i-1}$ is 
$-1/\sqrt{n}$ and $p_{i+1}$ is $0$ or vice versa.

$x_{i} = {2\sqrt{n} \over (c - 2m - 4l)}$ if both $p_{i-1}$ and $p_{i+1}$ 
are $1/\sqrt{n}$.

$x_{i} = 0$ if either $p_{i-1}$ is 
$1/\sqrt{n}$ and $p_{i+1}$ is $-1/\sqrt{n}$ or vice versa.

$x_{i} = {-2\sqrt{n} \over (c - 2m - 4l)}$ if both $p_{i-1}$ and $p_{i+1}$ 
are $-1/\sqrt{n}$.

For $p_{i} = 1/\sqrt{n}$:

$x_{i} = {n - m - 2l \over \sqrt{n}(c + m + 2l)}$ if both 
$p_{i-1}$ and $p_{i+1}$ are 0.

$x_{i} = {n - 2m - 4l \over 2 \sqrt{n}(c + m + 2l)}$ if either 
$p_{i-1}$ is $1/\sqrt{n}$ and $p_{i+1}$ is $0$ or vice versa.

$x_{i} = {3n - 2m - 4l \over 2 \sqrt{n}(c + m + 2l)}$ if either 
$p_{i-1}$ is $-1/\sqrt{n}$ and $p_{i+1}$ is $0$ or vice versa.

$x_{i} = { - m - 2l \over \sqrt{n}(c + m + 2l)}$ if both 
$p_{i-1}$ and $p_{i+1}$ are $1/\sqrt{n}$.

$x_{i} = {n - m - 2l \over \sqrt{n}(c + m + 2l)}$ if either $p_{i-1}$ is 
$1/\sqrt{n}$ and $p_{i+1}$ is $-1/\sqrt{n}$ or vice versa.

$x_{i} =  {2n - m - 2l \over \sqrt{n}(c + m + 2l)}$ if both $p_{i-1}$ 
and $p_{i+1}$ are $-1/\sqrt{n}$.

For $p_{i} = -1/\sqrt{n}$:

$x_{i} = -{n - m - 2l \over \sqrt{n}(c + m + 2l)}$ if both 
$p_{i-1}$ and $p_{i+1}$ are 0.

$x_{i} = -{3n - 2m - 4l \over 2 \sqrt{n}(c + m + 2l)}$ if either 
$p_{i-1}$ is $1/\sqrt{n}$ and $p_{i+1}$ is $0$ or vice versa.

$x_{i} = -{n - 2m - 4l \over 2 \sqrt{n}(c + m + 2l)}$ if either 
$p_{i-1}$ is $-1/\sqrt{n}$ and $p_{i+1}$ is $0$ or vice versa.

$x_{i} = -{2n - m - 2l \over \sqrt{n}(c + m + 2l)}$ if both 
$p_{i-1}$ and $p_{i+1}$ are $1/\sqrt{n}$.

$x_{i} = -{n - m - 2l \over \sqrt{n}(c + m + 2l)}$ if either $p_{i-1}$ is 
$1/\sqrt{n}$ and $p_{i+1}$ is $-1/\sqrt{n}$ or vice versa.

$x_{i} =  {m + 2l \over \sqrt{n}(c + m + 2l)}$ if both $p_{i-1}$ 
and $p_{i+1}$ are $-1/\sqrt{n}$.

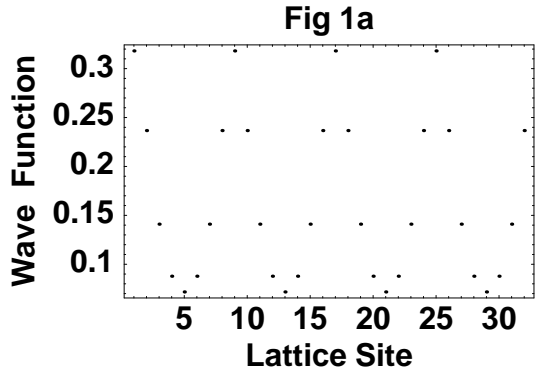
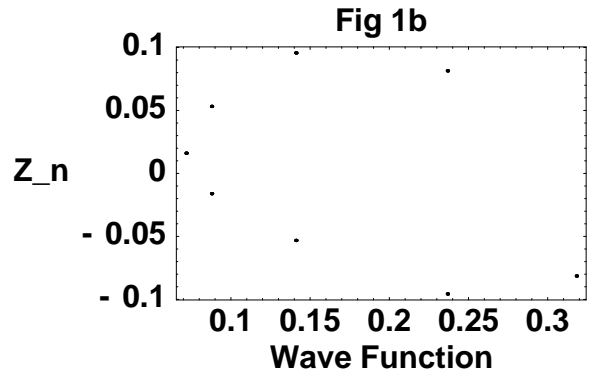
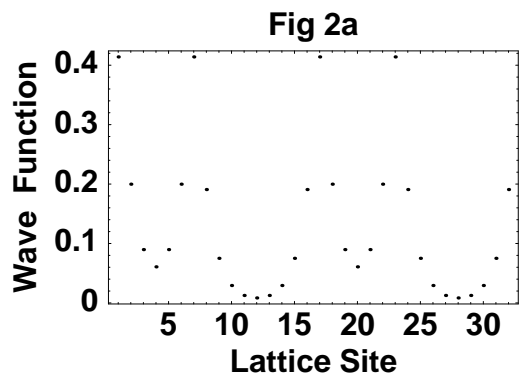
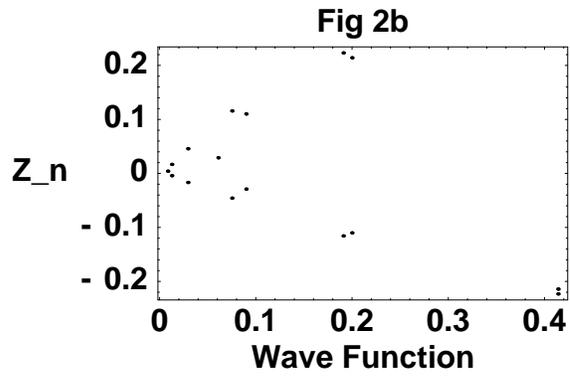
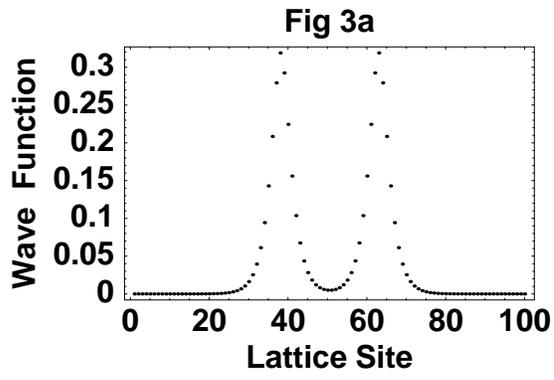
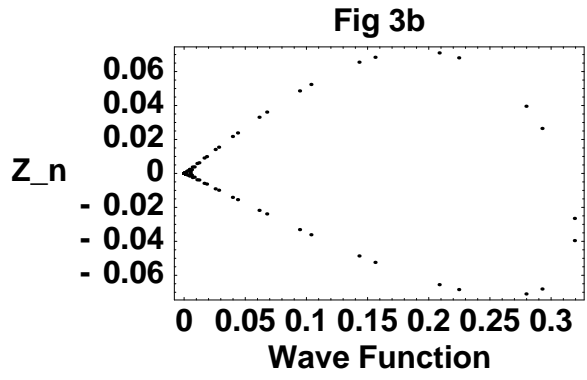
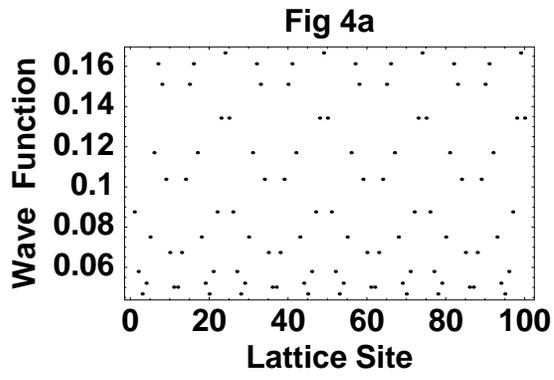
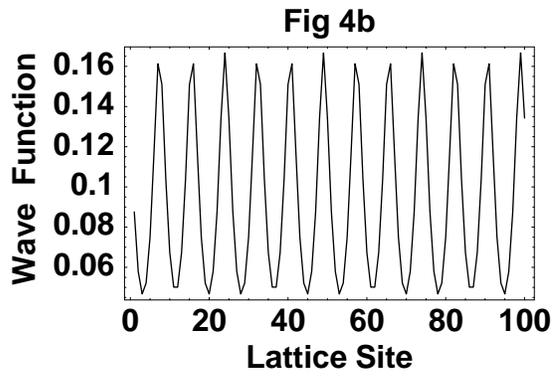
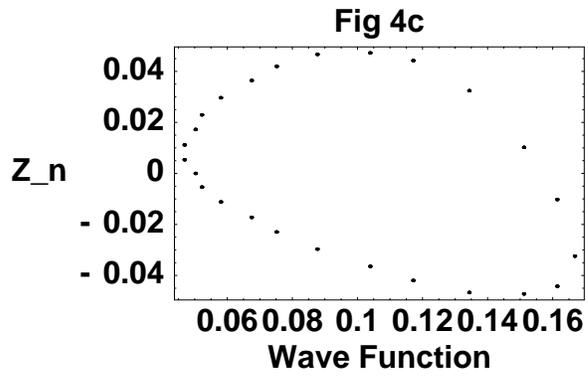
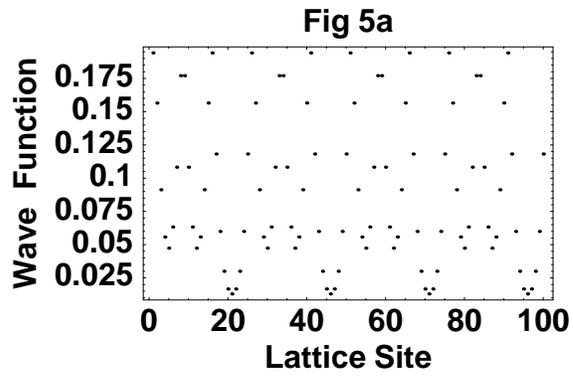
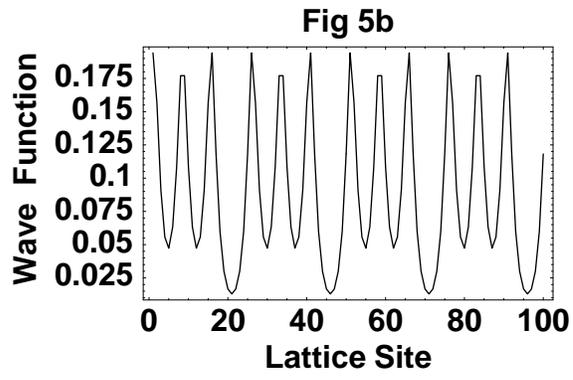
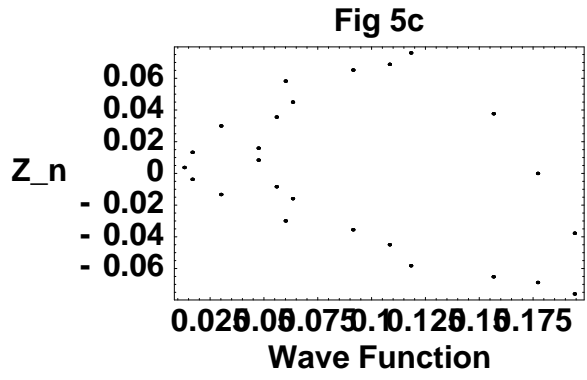
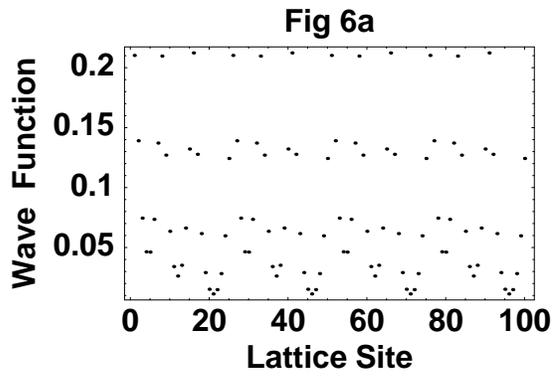
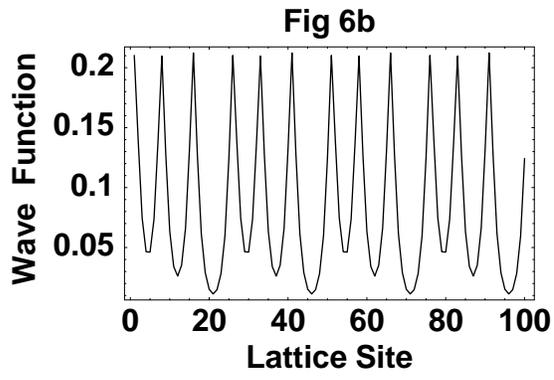
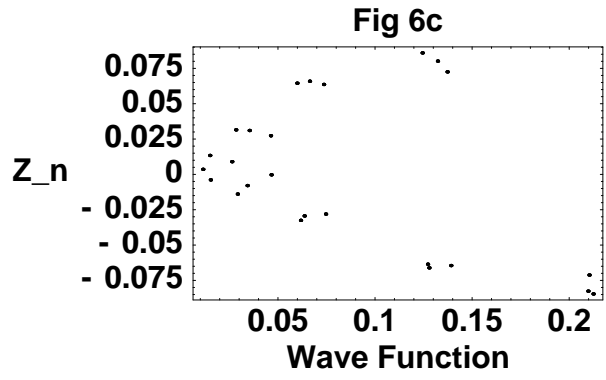
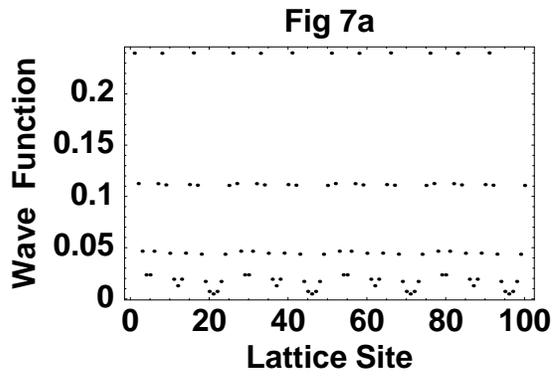
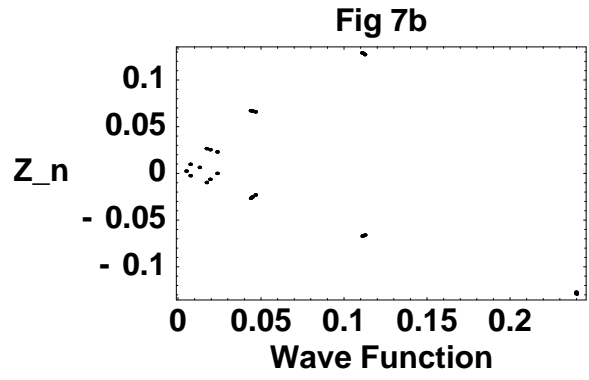
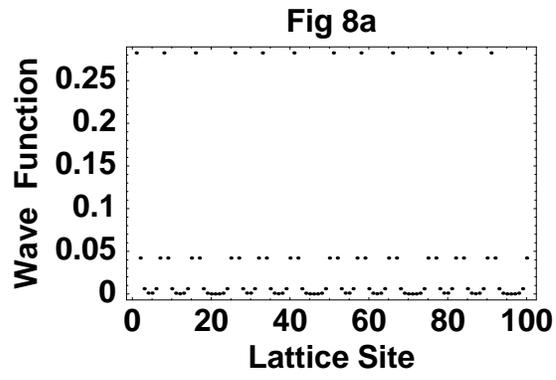
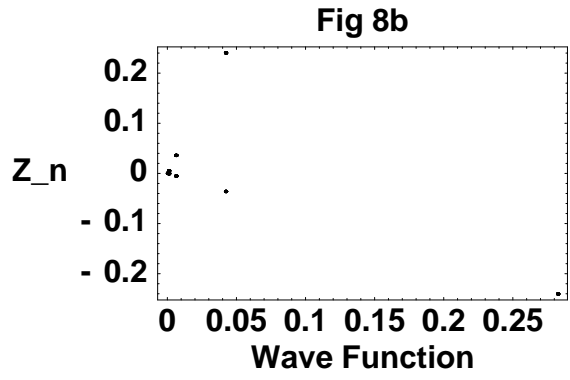
\end{document}